\documentclass[prl,twocolumn,showpacs,superscriptaddress]{revtex4-1}
\usepackage{amsfonts,amsmath,amssymb,graphicx}
\usepackage{amsfonts}
\usepackage{amsmath}
\usepackage{amssymb}
\usepackage{graphicx,color}
\newcommand{\f}{\frac}

\newcommand{\bbZ}{\mathbb{Z}}
\newcommand{\sgn}{\mathrm{sgn}}
\begin{document}

\title{Dynamical tunneling of a Bose-Einstein condensate in periodically driven systems}

\author{R. K. Shrestha,$^1$ J. Ni,$^1$ W. K. Lam,$^1$ G. S. Summy,}
\affiliation{Department of Physics, Oklahoma State University, Stillwater, Oklahoma 74078-3072, USA}
\author{S. Wimberger}
\affiliation{Institut f\"{u}r Theoretische Physik, Universit\"{a}t Heidelberg, Philosophenweg 19, 69120 Heidelberg, Germany}

\date{\today}

\begin{abstract}
\noindent
We report measurements of dynamical tunneling rates of a Bose-Einstein condensate across a barrier in classical phase space.
The atoms are initially prepared in quantum states which extend over a classically regular island region. We focus on the specific
system of quantum accelerator modes of the kicked rotor in the presence of gravity. Our experimental data is
supported by numerical simulations taking into account imperfections mainly from spontaneous emission. Furthermore we predict experimentally
accessible parameter ranges over which direct tunneling could be readily observed if spontaneous emission was further suppressed.
Altogether, we provide a proof-of-principle for the experimental accessibility of dynamical tunneling rates in periodically driven systems.
\end{abstract}

\maketitle

One of the first manifestations of quantum mechanics was radioactive decay, in which -- according
to Gamov's theory -- a particle can overcome a static potential barrier because of an even exponentially small
tail of its spatial wave function at the unbounded side of the barrier \cite{qm}. Besides this static
problem of over-the-barrier tunneling, dynamical tunneling mechanisms are also well known today \cite{heller}.
In classical dynamical systems, the phase space is generally mixed, with regions of regular and chaotic motion
\cite{LL92}. Though dynamical barriers forbid the classical transport between these regions, a quantum particle
can tunnel by leaking through classically invariant curves.

Dynamical or chaos-assisted tunneling has been subject of many theoretical works over the last years \cite{CAT,RAT,book}, which, in particular,
use tools of semiclassics to make approximate predictions about the quantum mechanical tunneling. Experiments to observe even signatures of
the tunneling of states initially prepared within classically regular regions into the chaotic surrounding, are notoriously hard \cite{exp,cold,kuhl}. The difficulties are
very natural since it is hard to control the effective Planck's constant $\hbar_{\rm eff}$ over a wide range. $\hbar_{\rm eff}$ determines the size of the quantum state
with respect to phase space. If $\hbar_{\rm eff}$ is large (see in particular \cite{cold}), tunneling occurs fast, but one is far from the
semiclassical limit which many fascinating theories exploit \cite{book}. On the other hand, if $\hbar_{\rm eff}$ is small, tunneling simply would take too long to be
measured precisely.

To balance this tradeoff we show the experimental realization of a paradigmatic system \cite{raizen} which not only provides
access to a wide range of $\hbar_{\rm eff}$, but also allows for precise control of the initial state and the dynamical evolution. The theoretical
description of our system and predictions on the resonance-assisted tunneling (RAT) mechanism can be found in \cite{shmuel}. We show a proof of principle
observation of dynamical tunneling experimentally, backed up by numerical simulations which also include the effect of experimental imperfections. Finally, we check the
scaling of the direct tunneling \cite {ketz} (i.e. not based on more complicated processes such as RAT) rates with the ratio of the area and the effective Planck's constant.
These predictions may be tested in future experiments once we overcome the main imperfection arising from spontaneous emission.

In our experiment we prepare a Bose-Einstein condensate (BEC) of about 40000 $^{87}$Rb atoms in the $5S_{1/2}$, $ F=1$ level using an
all-optical trap. After release from the trap, the BEC is exposed to kicks with period $T$ from a standing wave formed by two laser
beams of wavelength $\lambda=$ 780 nm, detuned $\Delta=2\pi \times 6.8 $GHz to the red of the atomic transition. The strength of these kicks is given by $k \approx \frac{\Omega^2}{\Delta} \Delta t$ where $\Omega$ is the Rabi frequency of the transition and $\Delta t$ is the duration of a pulse. The probability of a spontaneous emission event per atom during a kick can be calculated from $p_{\rm SE}=\frac{k}{\tau_{\rm SE}\Delta}$, where the
lifetime of the atomic state is $\tau_{\rm SE}=26 $ns. To control the phase, intensity, and pulse length as well as the
relative frequency between the kicking beams, each laser beam passes through an acousto-optic modulator driven by a waveform generator. Adding two counterpropagating waves differing in frequency by $\Delta f$ results
in a standing wave that moves with a velocity $v=2\pi\Delta f/G$. This fequency shift is used to accelerate the
lattice corresponding to an effective gravity field with dimensionless acceleration parameter $\eta=gMT/(\hbar G)$. $G=2\pi/\lambda_{G}$ is the
grating vector of the kicking lattice and $M$ the atomic mass. Additionally, the quasi-momentum of the BEC relative to the
standing wave is proportional to $v$, which is used to prepare the initial state within the relevant phase space region in momentum.
In dimenionless units (see \cite{review,fidelity,shmuel,italo}), momentum is decomposed as $p=n+\beta$ where $\beta$ $(0\leq\beta <1)$
is quasi-momentum and $n \in \bbZ$. The BEC is sufficiently dilute such that we can exclude interatomic interaction effects.

In this way we realize quantum accelerator modes as done previously in \cite{mode,gazal}. These
modes correspond to classically stable resonance islands embedded in a chaotic surrounding. They are perfect for our purpose
since the modes move with constant speed in momentum space, while the chaotic part essentially remains behind \cite{shmuel}. Hence, dynamical tunneling from
the islands to the surroundings is not hindered by backflow into the island.

\begin{figure}[t]
\includegraphics[width=0.9\linewidth]{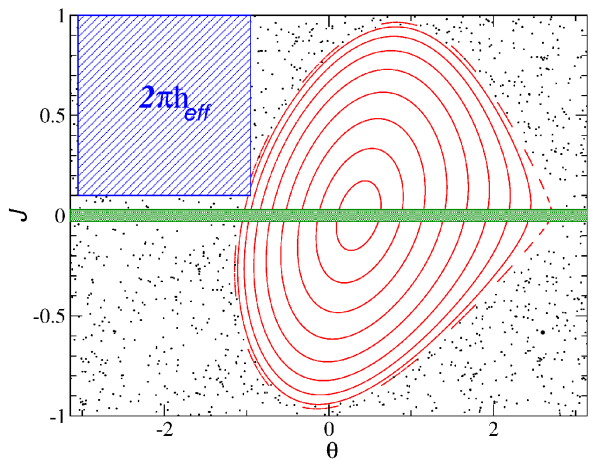}
\caption{(Color online) The phase-space of the pseudo-classical evolution given by Eq.~\eqref{eq:map} with $J$ taken $\textrm{ mod } 2\pi$. The initial state of the condensate (dashed green region) extends over the mid-point of the classical island (red). The effective Planck cell with $\hbar_{\rm eff}=|\epsilon|$ is shown by the (blue) shaded rectangle. For the given parameters, $\tau=5.97, \eta=0.0257, k=1.4$, more than half of the initial state follows the island motion and contributes to the tunneling signal.}
\label{fig:1}
\end{figure}

The dynamics of our kicked atom accelerator is described by the following Hamiltonian in the accelerated frame and in
dimensionless units \cite{italo,fidelity}:
\begin{equation}
\label{ham1}
\hat{H}(t)=\f{1}{2}\left(\hat{\mathcal{N}}+\beta+\eta\f{t}{\tau}\right)^2
+ k \cos(\hat{\theta})\sum_{j}^{}\delta (t-j).
\end{equation}
Here $\hat{\theta}=G\hat{x}\mod(2\pi)$, $\hat{\mathcal{N}} = -i\f{d}{d\theta}$ is the angular momentum operator, and $k$ and $\tau=T/T_{1/2}$
are the kicking strength and period. With periods close to $\tau=2\pi$, or $T$ close to the half-Talbot time $T_{1/2}=\f{2\pi M}{\hbar G^2}$,
the evolution of our quantum system is approximately dscribed by the following pseudo-classical map \cite{review,italo,shmuel}
 \begin{equation}
  \label{eq:map}
    \begin{array}{rcl}
      J_{j+1} &=& J_{j} + \tilde{k} \sin \theta_{j+1} + \sgn(\epsilon) \tau\eta \\
      \theta_{j+1} &=& \theta_{j} + \sgn(\epsilon)J_{j+1} \;\;\;\textrm{ mod } 2\pi \,.
      \end{array}
\end{equation}
Here $\tilde{k}=k|\epsilon|, \epsilon=\tau-2\pi$ and the classical momentum $J=n|\epsilon|+\sgn(\epsilon)[\pi+\tau (\beta+jn+\eta/2)]$.
We review such details since, by the latter formula, it is clear how control on the quantum numbers $n$ and $\beta$ allows us to prepare the initial state with a maximal
overlap with the classical stable regions \cite{gazal,sadgrove,fidelity}, see Fig.~\ref{fig:1}. In our experiments, $n$ is chosen zero and the
$\beta$ values obey approximately a Gaussian distribution with FWHM 0.06 centred at $\beta=0.5$.

\begin{figure}[t]
\includegraphics[width=1.05\linewidth]{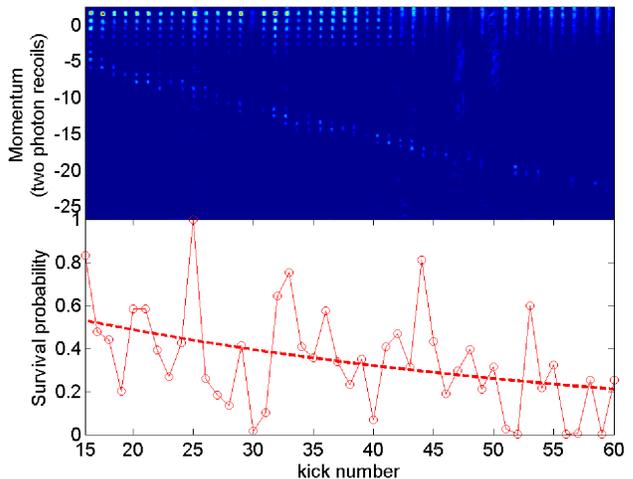}
\caption{(Color online) Upper panel: Experimental momentum distributions showing an accelerator mode in the free falling frame. Once the mode, composed of the
momentum states moving downwards, exits the bulk at $t \approx 15$, its population is simply measured by counting the number of atoms in it. Lower panel: Extracted
survival probability vs. kick number (solid line) and exponential fit (dashed line). Parameters are the same as in Fig.~\ref{fig:1}.}
\label{fig:2}
\end{figure}

\begin{figure}[t]
\includegraphics[width=0.95\linewidth]{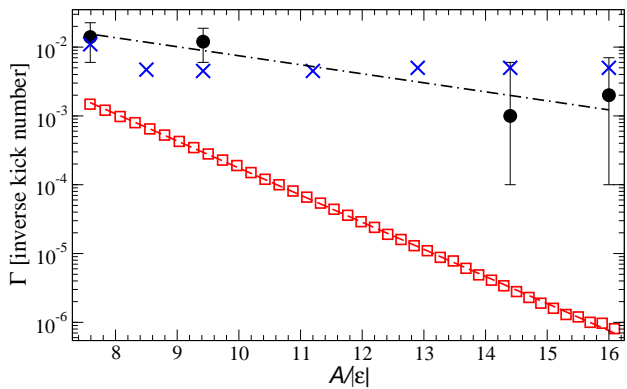}
\caption{(Color online) Decay rates of the accelerator modes (black circles) extracted from data such as shown Fig.~\ref{fig:2} and from numerical simulation for
long times without any imperfections (red squares) and with spontaneous emission (blue crosses). Parameters are the same as in the previous figures but with
$k=0.9, 1, 1.3,$ and $1.4$ for the experimental points. The global scaling with the area divided by Planck's constant
$A/|\epsilon|$ has a prefactor in the exponent of $\approx - 0.9$ (see fit by red dashed line) as compared to the theoretical value $-1$ for direct tunneling $\propto \exp(-A/|\epsilon|)$. The dot-dashed line presents a fit to the experimental data. There the slope is smaller, $\approx - 0.3$, originating from spontaneous emission which leads to the saturation of the curve with increasing $k$. The observation of a negative slope is strongly suggestive that dynamical tunneling is playing an important role in these decay rates. 
}
\label{fig:3}
\end{figure}

Our main observables are momentum distributions as a function of the number of standing wave pulses applied to the BEC. Figure~\ref{fig:2} (upper panel) shows such
a distribution for characteristic system parameters. In the lower panel we plot the relative amount of population in the mode vs. the kick counter. From this survival
probability (to stay within the mode or within the classical resonance island) we extract the decay rates by exponential fits. These rates are presented
in Fig.~\ref{fig:3} together with decay rates extracted from numerical simulations. We see that the experimental rates are systematically larger than the
ideal numerical ones. Hence we run simulations also including random events of spontaneous emission occurring with a probability of $p_{\rm SE}=5\times10^{-3}k$ per single kick and per atom.
Within the experimental error bars, we then obtain reasonable agreement with the experimental data. One notable feature of the experimental data is that there is a distinct reduction in the decay rate as area increases. This is exactly the opposite of the behavior expected if our decay rates are primarily due to SE since $k$ and hence $p_{\rm SE}$ increase in this situation. While there is also the possibility of other types of noise such as vibrations of the lattice, this only becomes significant as the duration (i.e. number of kicks) of the experiment is extended. It is for this reason that we could only measure up to a relatively small number of kicks of about $t \approx 60$. Hence the temporal range to extract the experimental rates is rather small.

Please note that we plotted the rates in Fig.~\ref{fig:3} as a function of the area $A$ of the resonance island shown, e.g., in Fig.~\ref{fig:1}. These
areas are divided by the effective Planck's constant $|\epsilon|$. We computed the areas for different parameter sets by propagating a single classical
trajectory in the chaotic region for many kicks, see appendix B in \cite{remy} for details. A semiclassical theory for direct tunneling (i.e. without resolving additional small-scale structures within the island) would predict a scaling $\Gamma \propto \exp(-A/|\epsilon|)$. Interestingly, the ideal numerical simulations show a slightly smaller slope of $\approx -0.9$. The simulations are done for ensembles of $10^4$ $\beta$-rotors with a Gaussian distribution of the $\beta$ values as specified above. The decay rates of the accelator modes (red squares in Fig.~\ref{fig:3}) are extracted from summing up the contribution of five to ten momentum states in the modes (as seen in the upper panel of Fig.~\ref{fig:2}) as a function of time for up to a maximum of $5 \times 10^4$ kicks. Since the corresponding computations with spontaneous emission take a long time we restrict to just a few data points, shown by the blue crosses, in this case.

In Figure~\ref{fig:3}, we observe clear deviations from the scaling $\Gamma \propto \exp(-A/|\epsilon|)$ for the experimental data arising from the saturation of the curve with increasing kick strength $k$ or increasing $A/|\epsilon|$, respectively. This is due to spontaneous emission which scales linear in $k$, i.e. $p_{\rm SE} \propto k$. In consequence, spontaneous emission really hinders us to observe the expected scaling for the direct tunneling rates. Nevertheless as noted previously, the experimental decay rate does decrease with area, which strongly suggests that dynamical tunneling is playing an important role. Thus even though there are limitations to our experimental study, it can still be seen as a demonstration of the proof-of-principle for the observation of dynamical tunneling in periodically kicked systems with classical mixed phase space.

\begin{figure}[t]
\includegraphics[width=\linewidth]{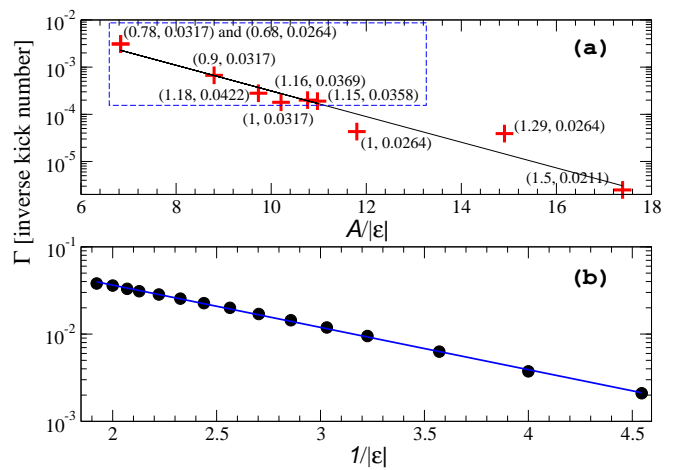}
\caption{Numerical simulations: (a) dynamical tunneling rates for fixed  $\tau=5.8$ and the shown tuples of $(k,\eta)$ vs.
the area $A$ of the resonance island of the modes over Planck's constant $|\epsilon|$. (b) tunneling rates for fixed $A$ (fixed classical phase space structure) with $\eta=0.06$
changing $k$ and $|\epsilon|$ accordingly such that $k|\epsilon|=0.5$. The solid lines present exponential fits with prefactors in the exponents $\approx -0.63$ in (a) and
$\approx -1.1$ in (b).}
\label{fig:4}
\end{figure}

To further check the semiclassical prediction for the direct tunneling process from the island into the chaotic sea, i.e. without any complications due to RAT
or other more elaborated processes, we searched for realistic parameter sets summarized in Fig.~\ref{fig:4}. In (a) we kept the kicking period $\tau=5.8$ fixed
and chose tuples of $(k,\eta)$, with $0.68 \leq k \leq 1.5$ and $0.0211 \leq \eta \leq 0.0422$, as indicated in the figure. In (b) we kept fixed the classical
phase space structure determined solely by the two parameters $\tilde{k}=k|\epsilon|=0.5$ and $\eta=0.06$. In both cases we observe an overall scaling with
$\Gamma \propto \exp(-A/|\epsilon|)$, while $A$ remains constant in (b) in contrast to (a). In our imaging system, we cannot use higher $\eta$ and $k$ because the momentum distribution is not observable after a larger kick number. For instance, already with $\eta\approx 0.03$ and $k \approx 1.5$, the momentum distribution beyond 65 kicks cannot be observed  (see Fig. \ref{fig:2} (upper panel)).

Despite the experimental imperfections hindering us from measuring small tunneling rates precisely, the fact that the decay rate decreases with larger amounts of spontaneous emission shows that we have a proof-of-principle for the detection of dynamical tunneling in a periodically kicked Bose-Einstein condensate. If it were possible to reduce the spontaneous emission below $p_{\rm SE} < 10^{-4}$, the tunneling rates in Fig.~\ref{fig:4}(a) which lie in the range $2 \times 10^{-4}\ldots 4 \times 10^{-2}$ (inside the dashed box), should be experimentally measurable with our setup. This could be achieved with the help of a laser with a larger detuning from the atomic transition and a correspondingly higher intensity (to maintain the potential strength we currently have). Detecting more structure in the rates, e.g. arising from RAT \cite{book,shmuel}, is a harder task with an atom-optics setup. One possible solution might be to realize accelerator modes in an optical system where experiments can be carried out up to a few thousand kicks \cite{peschel}.  This would allow for better measurements of dynamical tunneling over a much larger range of the effective Planck's constant $|\epsilon|$.

\section*{Acknowledgements}

This work was partially supported by the NSF under Grant No. PHY-0653494 and by the DFG through the Research Unit FOR760 (WI 3426/3-1), the HGSFP (GSC 129/1), the Center for Quantum Dynamics and the Enable Fund of Heidelberg University. SW is grateful to R\'emy Dubertrand for his help in the preliminary stage, to Carlos Parra-Murillo and Roland Ketzmerick for useful comments, and for the cordial hospitality at OSU.

\end{document}